\documentclass[prl,aps,10pt,groupedaddress,amsmath]{revtex4}
\usepackage{graphicx} %For preprint/review copy only.
\usepackage{bm}% bold math
\usepackage{dcolumn}% Align table columns on decimal point

\def\iti {{\it i}} \def\itj {{\it j}}

\begin{document}

\title{Origin of Diffuse Scattering in Relaxor Ferroelectrics}
\author{P. Ganesh $^{1}$, E. Cockayne $^{2}$, M. Ahart $^{1}$, R. E. Cohen $^{1}$, B. Burton $^{2}$,
        R. J. Hemley $^{1}$, Yang Ren $^{3}$, Wenge Yang $^{3}$, Z-G Ye $^{4}$}
\affiliation{$^1$ Geophysical Laboratory, Carnegie Institution of Washington, 5251 Broad Branch Road, Washington, DC 20015, USA\\
$^2$ Ceramics Division, Materials Science and Engineering Laboratory, National Institute of Standards and Technology, Gaithersburg, MD 20899 USA\\
$^3$ Advanced Photon Source, Argonne National Laboratory, Argonne, Illinois 60439, USA\\
$^4$ Department of Chemistry, Simon Fraser University, Burnaby, BC, V5A 1S6, Canada
}

\date{\today}

\begin{abstract}

High-pressure and variable temperature single crystal synchrotron
X-ray measurements combined with first-principles based molecular
dynamics simulations study diffuse scattering in the relaxor
ferroelectric system PSN (PbSc$_{1/2}$Nb$_{1/2}$O$_3$).  Constant
temperature experiments show pressure induced transition to the
relaxor phase at different temperatures characterized by butterfly and
rod shaped diffuse scattering around the $\{$h00$\}$ and $\{$hh0$\}$
Bragg spots, respectively.  The simulations reproduce the observed
diffuse scattering features as well as their pressure-temperature
behavior, and show that they arise from polarization correlations
between chemically-ordered regions, which in previous simulations were
shown to behave as polar nanoregions.  Simulations also exhibit radial
diffuse scattering (elongated towards and away from {\bf Q}=(000)),
that persists even in the paraelectric phase, consistent with previous
neutron experiments on (PbMg$_{1/3}$Nb$_{2/3}$O$_3$) (PMN).  

%PG: Ron wanted this line removed from abstract

%These features arise from Coulomb forces of the Sc$^{+3}$ and
%Nb$^{+5}$ ions in the chemically disordered regions.

%OFF-SUBJECT By engineering COR sizes and
%distribution it should be possible to control relaxor behavior,
%allowing greater control over actuator and tunable dielectric
%applications.

\end{abstract}

\maketitle 
%%\section{\label{sec:intro}Introduction}

Single crystal relaxors have huge electromechanical coupling, and show
much promise for ultrasonic transducer
applications~\cite{ParkShrout}. They have broad frequency
and temperature dependent dielectric maxima, which in the special case
of relaxor-ferroelectrics~\cite{Samara1,Samara2} drops to a much lower
value below the ferroelectric transition temperature ($T_{FE}$).  The
origin of the relaxor phase has been a topic of intense research for
over a decade.  From refractive index measurements, Burns et
al.~\cite{Burns} suggested that formation of polar clusters below a
characteristic temperature, now called the Burns temperature
($T_d$)~\cite{Burns}, gives rise to dielectric dispersion, which was
confirmed experimentally~\cite{Jeong}.  These clusters thought to be a
few nanometers in size are called polar nanoregions (PNRs).

Recent X-ray and neutron experiments show characteristic shapes of
diffuse scattering in the relaxor phase of several lead based
relaxors~\cite{GXuPcake,MuhtarPZNPT,PierrePZN,PGehring2009}
absent in their paraelectric or ferroelectric phases.  The main
observed feature is anisotropic diffuse scattering around the Bragg
peaks along $<$110$>$ directions.  The ferroelectric phase shows weak
streaks similar to BaTiO$_3$ and KNbO$_3$, but rotated by 45-degrees.
The paraelectric phase only shows radial diffuse scattering.
%EC: I would suggest leaving out the follwing for length.
%They have been included in the fig. caption
%Focusing on the (hk0) plane 
%((hkl) indicates Miller planes, {hkl} the
%set of symmetry related (hkl) planes, [hkl] indicates directions and
%$<$hkl$>$ indicates the set of symmetry equivalent directions), 
%we refer to the enhanced intensity at (hh0) Bragg peaks along
%[1$\bar{1}$0] as ``rods", and the enhanced intensity at (h00) peaks
%along both [110] and [1$\bar{1}$0] directions as ``butterflies".  In
%addition, ``radial'' diffuse scattering (directed towards and away from
%{\bf Q}=(000)) is observed at all the Bragg peaks, which remain even
%in the paraelectric phase.

%PG: The sentence below makes it too obvious that PNRs cause DS, main result of our paper! So Ron suggested removing it
%Because polar nanoregions have enhanced local polarization relative to
%the global polarization, and because polarization in perovskite oxides
%is dominated by ionic displacements, PNR must be somehow associated
%with diffuse scattering observable in neutron and X-ray diffraction
%experiments.  
Several hypotheses previously advanced to explain the characteristic
shapes ~\cite{GXuPcake,Pasciak,Vakhrushev,Welberry}, invoke some type
of artificial anisotropic features such as anisotropic strain,
correlated atomic shifts, or domain boundaries to generate the
experimentally observed anisotropic diffuse scattering features.
Fitting the shapes of diffuse scattering features is not
sufficient to uniquely determine the nature of the microstructural
feature that give rise to them.
%Furthermore, none of the above models
%incorporate realistic polarization fluctuation physics; rather
%anisotropic features are artificially imposed.  
Incorporating realistic polarization fluctuations via first-principles
derived models provide a basis for clarifying the nature of the PNR in
relaxors and experimental diffuse scattering observations
provide a critical test for any theoretical model.  In this Letter, we
clarify the microstructural origin of diffuse scattering features in
the relaxor- and the paraelectric- phases of PSN, and related
materials, by combining single crystal X-ray diffraction experiments
and first-principles-based molecular-dynamics
simulations~\cite{Tinte}.  Our result suggests new pathways to
engineer novel materials with superior electro-mechanical properties.

%and understand how they
%are related to relaxor behavior and the large
%electro-mechanical coupling observed in relaxor materials.

\begin{figure}
\includegraphics[width=5.5in,angle=-90]{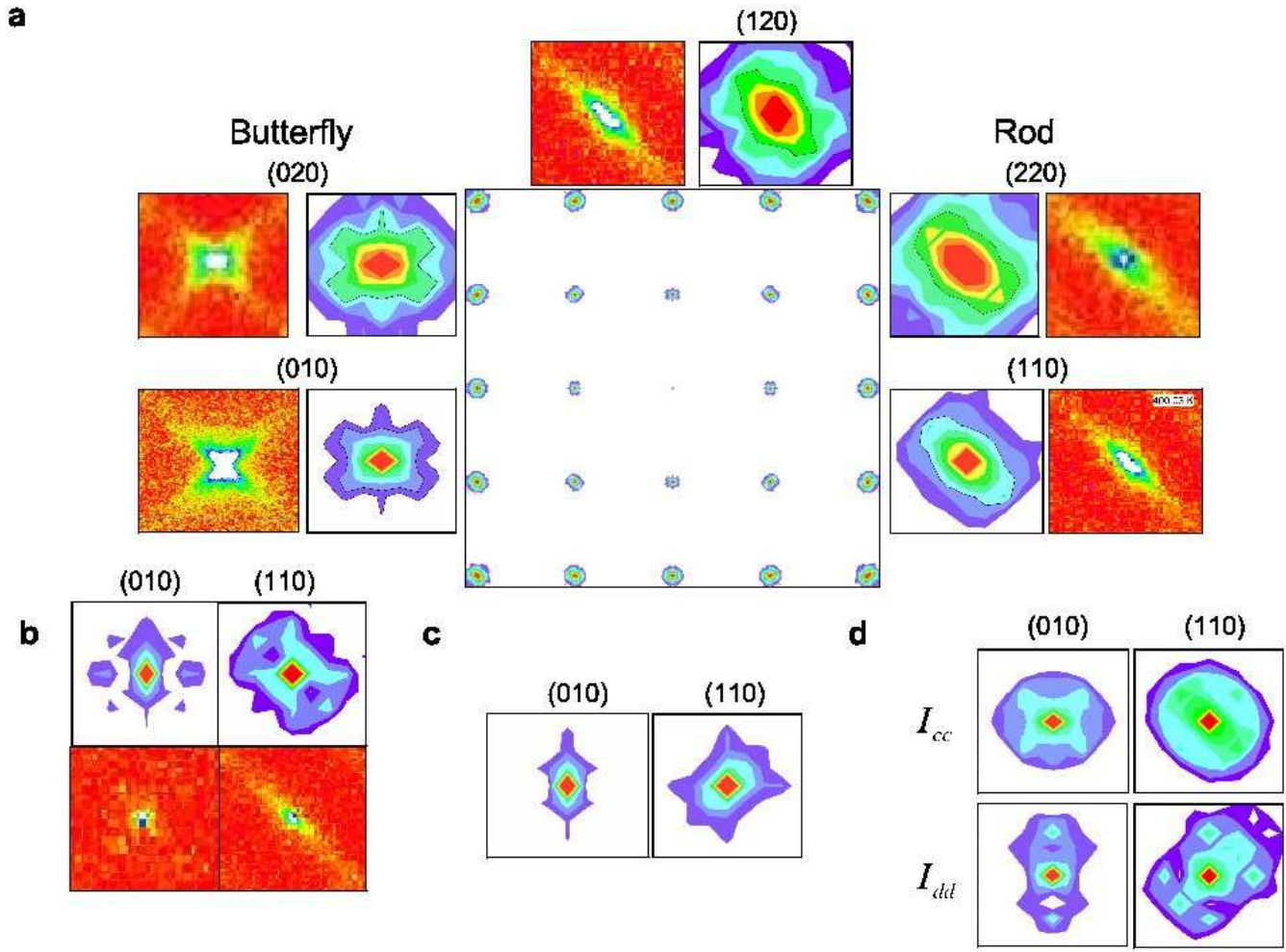}
\caption{\label{fig:fig1} Temperature dependence of diffuse scattering
in (hk0) plane from X-ray diffraction at ambient pressure (red
background) and simulations at P=18GPa (white background) (a)
Characteristic butterfly and rod shaped diffuse scattering (enhanced
intensity along $[$110$]$ and $[1\bar{1}0]$ directions avoiding {\bf
Q} =(000)) is observed in the relaxor phase around $\{$h00$\}$ and
$\{$hh0$\}$ Bragg spots, respectively, in experiments (400~K) and
simulations (200~K) (b) In the ferroelectric phase, the characteristic
shapes diminish in experiments (250~K), similar to simulations (10~K)
(c) In the paraelectric phase, simulations (320~K) only show radial
diffuse scattering (d) $I_{cc}$ only shows the butterfly and rod
shapes, while $I_{dd}$ shows only the radial patterns in the simulated
relaxor phase.}
\end{figure}

%%\section{\label{sec:Exp}Experiment}

X-ray diffraction was measured on single crystal PSN at beamlines
11-ID-C (0.1077~\AA) and 16-BMD (0.436926~\AA) of the Advanced Photon
Sources (Argonne National Laboratory).  Single crystals of disordered
PSN were grown from high-temperature solutions using a mixture of
PbO and B$_2$O$_3$ as solvent~\cite{Bing}. A crystal with dimension of
%EC: Describe thermal processing of the PSN sample
%PG: Not sure what you mean here.  Do you mean the thermal history of the sample?
%EC: Annealing temperature/atmosphere?  The amount of ordering in PSN can
%EC controlled by heat treatment, and the amount of Pb vacancies
%EC depends on the atmosphere in which the sample was made.
%PG: The exact details are not known, but it is supposedly
%``disordered PSN'' and remains so throughout exp. since the higheest
%temp. in exp. is close to T_Burns 
70x70x20 $\mu$m$^3$ and oriented along (001) was loaded into a diamond
anvil cell (DAC) with Ne as the pressure medium. A ruby chip and a
small grain of Au were also loaded for pressure determination
(accuracy of 0.2 GPa).  A MAR3450 image plate was used to record
oscillation photographs.  For low temperature measurements, the DAC
was loaded in a He flown cryostat, in which temperature was measured
with a thermocouple with an accuracy of $\pm$ 2 K.  Additional
experimental details about beamline 11-ID-C and 16-BMD can be found in
Refs.~\cite{Rutt,YFeng}.  To facilitate measurement of the diffuse
scattering signal, the sample was rocked $\pm$ 6$^\circ$ in omega.
Because of its (001) orientation in a cryostat, only the (hk0) indexed
Bragg peaks could be observed.

%PG: These details were commented due to lack of space.
%Single crystal diffraction patterns were first measured at ambient
%pressure at various temperatures between 250 and 400~K. Temperature
%was subsequently decreased, and at each temperature of T=300~K, 130~K
%and 50~K, pressure was applied and X-ray data was collected.  Because
%the single crystal sample was loaded into the the DAC in the cryostat
%with a (001) orientation, only the (hk0) indexed Bragg peaks could be
%observed.  
At ambient pressure, PSN enters the relaxor phase below the Burns
temperature (~T$_d$ $\sim$ 650~K) and undergoes a ferroelectric
transition at T$_{FE}$ $\sim$ 365~K. In our experiments at 400~K
(Fig.~1{\bf a}), we observe the characteristic butterfly and rod
shaped diffuse scattering.  At 250 K (Fig.~1{\bf b}), below T$_{FE}$,
the characteristic diffuse scattering diminishes considerably.  Diffuse
streaks connecting the Bragg peaks grow weaker, yet persist in the
ferroelectric phase, similar to those observed in BaTiO$_3$ and
KNbO$_3$ (Ref.~\cite{BTKNbDS} and references therein).

We observe pressure induced phase transitions at 300~K, where around
1.4GPa a reappearance of the relaxor phase characterized by diffuse
scattering is seen; persisting up to 1.5GPa, but disappearing at
1.8GPa (Fig.~2{\bf b}). Disappearance of diffuse scattering at
higher-pressures indicate a ferroelectric to paraelectric phase
transition consistent with dielectric-measurements~\cite{Samara1}.  

%At cryogenic temperatures of 50~K (Fig.~2{\bf b}), we observe similar
%pressure dependence.

At 50~K (Fig.~2{\bf b}) we also observe superlattice peaks at low
pressures of P=0.6~GPa at (h+$1/2$ k+$1/2$ 0) suggesting a lowering of
symmetry, which reduce in intensity as pressure drives the system to
the relaxor phase. Our experiments are on disordered PSN, hence the
peaks are not due to ordering~\cite{RandallBhalla} but due to possible
octahedral rotations. In the relaxor phase, at 2.7~GPa, additional
satellite peaks with diffuse wings emerge, which persist up to 10~GPa.
%Presence of defects incommensurate with the symmetry of
%the lattice gives rise to satellite spots~\cite{SSpeak}.  
Ordering PNRs with a range of closely spaced incommensurate modulation
vectors will give similar satellite spots with diffuse
wings.

We performed molecular-dynamics simulations using a model Hamiltonian
obtained by expanding the potential energy of PSN about a high
symmetry perovskite reference structure and projecting onto the
subspace of soft normal modes, dominated by Pb displacements,
including the ferroelectric instabilities~\cite{Rabe2,Rabe3}.
%In PSN, these modes are dominated by displacements of Pb atoms, and
%only Pb-modes are explicitly included in the model.  
Model parameters are fitted to first-principles density functional
calculations~\cite{BurtonPSN,Notes}.

The model has chemically ordered regions (COR) with rock-salt ordering
of ``Sc$^{+3}$'' and ``Nb$^{+5}$'' 'B' site ions, embedded in a
chemically disordered region (CDR) that has a random 'B' site
configuration. This gives rise to a quenched ``random" component to
the local electric fields at the Pb-sites.  Relative to equal volumes
of CDR, polarization is enhanced in CORs, thereby acting as PNRs.  The
CORs are spatially fixed, because chemical order (disorder) is
quenched, but their average polar orientations vary
dynamically~\cite{BurtonPSN}. The model allows homogeneous strain to
fluctuate; inhomogeneous strain, higher-frequency contributions to
lattice polarization, and oxygen octahedral tilting are ignored.

%Simulations on the PSN model exhibit ferroelectric, relaxor and
%paraelectric phases depending on the temperature and
%pressure~\cite{Tinte}.  These phases are identified by the
%polarization in the COR.  Simulated polarization is enhanced in CORs,
%relative to equal volumes of CDM (chemically disordered regions, CDR).
%The COR are therefore nanoscale regions that exhibit enhanced
%polarization, PNR, below a simulation temperature identified with the
%Burns temperature.  The COR are spatially fixed, because chemical
%order (disorder) is quenched, but their average polar orientations
%vary dynamically ~\cite{BurtonPSN}.

%The 20 COR are assumed to be approximately spherical, and their positions
%within the the CDM were chosen randomly.  

%EC:Details of MD; type of algorithm (if any), timestep, thermostat, etc?
%PG: details are incuded below

%PUT THESE IN NOTES
%Molecular dynamics simulations were performed for durations of about
%300~ps at several temperatures and pressures spanning the
%ferroelectric, relaxor and the paraelectric phase fields.  A MD
%time-step of 0.6 fs was chosen.  A Nose thermostat was used to fix the
%temperature and a barostat to fix the pressure (hydrostatic pressure)
%that only allowed for homogeneous volume changes.  Scattering
%intensities were calculated only after initial equilibration of $\sim$
%150~ps.  

In X-ray and neutron scattering experiments, the measured
intensity is a Fourier transform of the ensemble averaged two point
density-density correlation.  Assuming point particles,
the intensity is:
%\begin{eqnarray}
%\label{eq:int1}
$I(\bf Q)=<\sum_{i,j}\it {e}^{\bf iQ \cdot (\bf
{r}_\iti-\bf{r}_\itj)}>=<|\sum_i \it {e}^{\bf {iQ} \cdot \bf {r}_\iti}
|^{2}>$
%\label{eq:int2}
%\end{eqnarray}
where $i$ runs over the total number of Pb-atoms in the supercell (Pb
scattering is expected to dominate our experiments as it
has a disproportionately large X-ray scattering factor) and $\bf{Q}$
values commensurate with the supercell were chosen.  Because our
distribution of COR in our supercell is not entirely isotropic, we
apply full cubic symmetry to the scattering intensity before comparing
with experiments.

\begin{figure}[htp]
\includegraphics[width=5.5in,angle=-90]{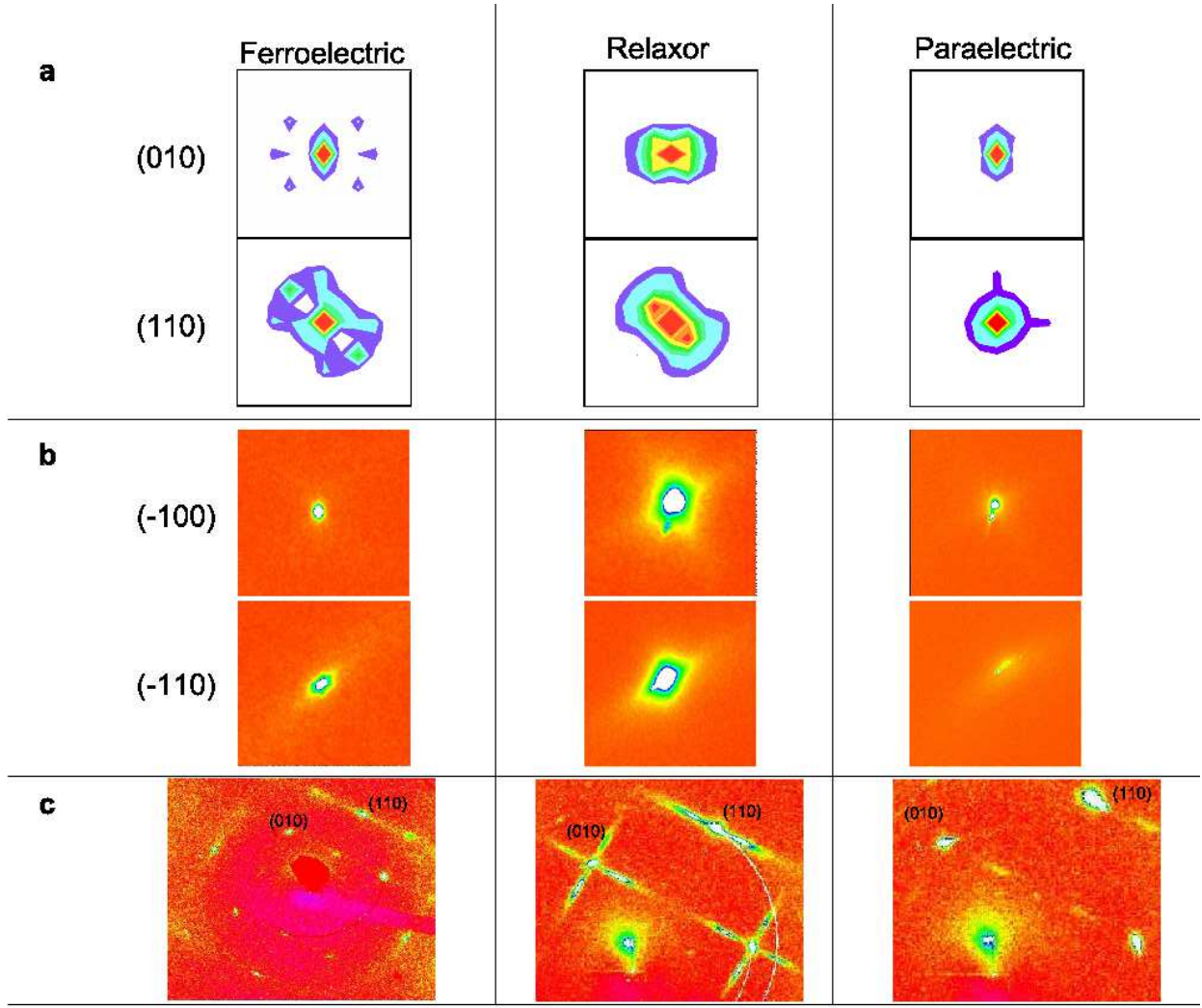}
\caption{\label{fig:fig2} Pressure dependence of diffuse scattering in
  simulations at 180~K (white background) and experiments (red
  background). (a) Simulated diffuse scattering in the ferroelectric
  (P=14~GPa), relaxor (P=18~GPa) and the paraelectric (P=25~GPa)
  phases respectively, show characteristic relaxor-phase butterfly and
  rod shaped diffuse scatting and radial diffuse scattering in the
  pressure induced paraelectric phase (b) Experiments at 300~K show
  weak diffuse scattering in the ferroelectric phase (P=0~GPa) but
  strong butterfly and rod shaped scattering in the relaxor phase
  (P=1.4~GPa).  In the paraelectric phase (P=1.8~GPa) the butterfly
  and rod shapes disappear. (c) At 50~K, we observe superlattice peaks
  in the ferroelectric phase (P=0.6~GPa).  In the relaxor phase
  (P=2.7~GPa) on top of the butterfly and rod shapes we observe
  satellite spots.  Further increase in pressure to 10~GPa destroys
  the diffuse scattering, leading to a pressure induced relaxor to
  paraelectric phase transition.  (The bright spot at the center of
  the image is due to the beam stopper.)}
\end{figure}

\begin{figure}
\includegraphics[width=5.5in,angle=-90]{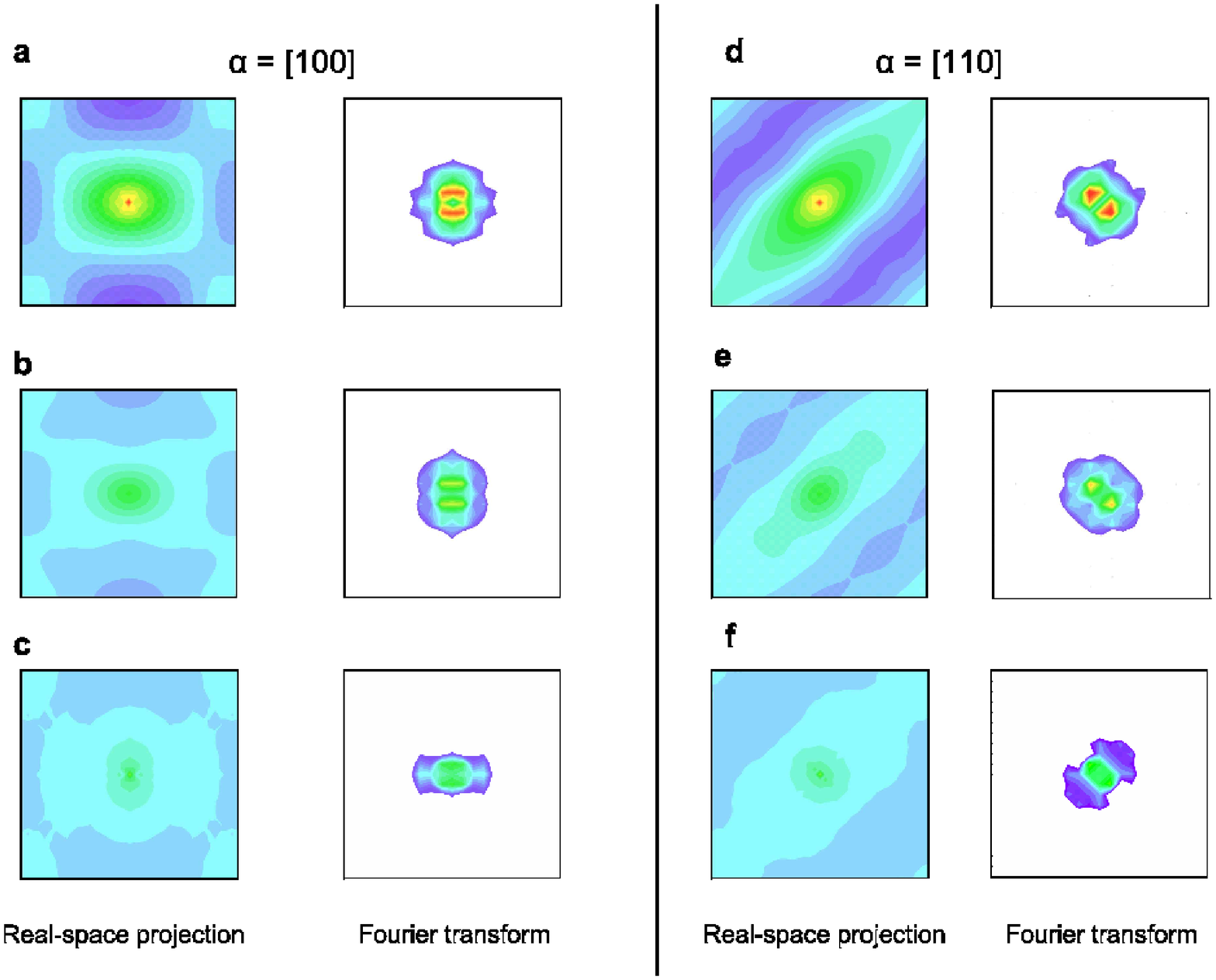}
\caption{\label{fig:fig3} Real-space correlations
(Eq.~\ref{eq:realcorr}) projected along [100] and [110] directions and
% Are we not using [100] and [110] for directions and 
% (100), (110) for planes?  Throughout the manusctipt,
% we use <hkl>, (hkl) and [hkl] for <spots>, (planes), and 
% [directions] (respectively?).  Are we being entirely consistent?
% Would it not be better to use the same brackets for spots and planes?
%EC: Also, in many of the figures, it seems that 100 is used when 
%010 is meant or vice versa.  Be careful about the labeling of peaks!
their direct Fourier transforms. (a) and (d) show correlations between
all 'Pb' atoms, which reproduce the experimentally observed butterfly-
and rod- shaped diffuse scattering. (b) and (e) show the correlations
between COR-COR regions which are extended in the real space along
directions normal to the butterfly and rod directions.  Their Fourier
transform gives the characteristic diffuse scattering. (c) and (f)
show the correlations between the CDR-CDR regions which when Fourier
transformed give the radial patterns}.
\end{figure}

Fig.~1 shows the temperature dependence of our computed diffuse
pattern in the (hk0) plane for the ferroelectric, relaxor and the
paraelectric phases at P=18~GPa.  In the relaxor phase~(Fig.~1{\bf a})
we observe the characteristic butterfly and rod shaped diffuse
scattering around (h00) and (hh0) Bragg spots, which diminish in their
intensity in the ferroelectric phase~(Fig.~1{\bf b}) in excellent
comparison to experiments.

Pressure dependent changes in diffuse scattering (Fig.~2{\bf a}) are
similar to those induced by varying temperature. Again, our model
captures the experimental pressure induced phase-transition at 300K
(Fig.~2{\bf b}). Experimentally observed streaks, like in KNbO$_3$
have been shown to be due to hopping between equivalent
sites~\cite{Krakauer}, and superstructure peaks at 50~K are possibly
due to octahedral rotations, both of which are excluded from our
model.  But these atomic features do not cause the relaxor phase,
which is characterized by the butterfly and rod shaped scattering,
which our simple model is able to capture.

As the system is driven towards the paraelectric phase, by increasing
temperature or pressure, the characteristic butterfly and rod shapes
vanish; however, a weak diffuse pattern that extend radially towards
and away from the origin ({\bf Q}=(000)) persists around all the Bragg
peaks.  Its intensity is larger in the direction away from the center
than towards it, consistent with recent neutron
experiments~\cite{PGehring2009} but absent in X-ray due to
their low intensity.
% We do not observe these features in
%our X-ray experiments, perhaps due to their very low intensity.  

%EC: Any evidence that sufficient simulation pressure causes
%the diffuse scattering to vanish again, as experimentally
%shown?
%PG: Yes, the paraelectric phase obtained by increasing the pressure
%leads to disappearance of rods and butterfly, with only the radial DS
%remaining.

We find no significant evolution of the diffraction patterns during
our simulations, which access frequencies greater than 10GHz.  This
suggests only static
%PNR-PNR
correlations cause diffuse scattering, consistent
with recent experiments on PMN~\cite{PGehring2009}.

We find the contributions to the scattering intensity from
correlations due to atoms in the different regions (COR and CDR).
% This gives,
%\begin{equation}
%\label{eq:Icc}
% I_{cc}({\bf Q})=<|\sum_{\iti \in CDR}{\it e}^{\bf iQ \cdot {\bf R}_{0 \iti}}+\sum_{\iti \in COR} {\it e}^{ \bf iQ \cdot {\bf r}_\iti}|^2>
%\end{equation}
%and
%\begin{equation}
%\label{eq:Idd}
% I_{dd}({\bf Q})=<|\sum_{\iti \in COR}{\it e}^{\bf iQ \cdot {\bf R}_{0 \iti}}+\sum_{\iti \in CDR} {\it e}^{ \bf iQ \cdot {\bf r}_\iti}|^2>
%\end{equation}
%where $\bf R_{0 \it i}$ is the undistorted position of Pb at site $i$.
These are shown in Fig~1{\bf d}, where $I_{cc}$ and $I_{dd}$ are the
intensities obtained by setting CDR and COR Pb displacements to zero
respectively.  While $I_{cc}$ shows characteristic rod and butterfly
features, the radial diffuse pattern is absent.  $I_{dd}$ only shows
weak radial diffuse pattern.
%The plot of $I_{dd}$ is
%similar to the full intensity computed for a different simulation with
%a random `Sc$^{3+}$':`Nb$^{5+}$' distribution (not shown).  
Thus, it is the correlation between COR atoms that give rise to the
characteristic diffuse pattern.  The CDR Pb-displacements are
dominated by strong electric fields fixed in a ``quenched''
distribution of ``Sc'' and ``Nb'' atoms in the chemically disordered
matrix, and therefore radial scattering persists even in the
paraelectric phase.  This is consistent with experimental observations
on PMN~\cite{PGehring2009}.  Note that our interpretation that radial
diffuse scattering is associated with chemical disorder differs from
the interpretation of Gehring {\it et al.} that it indicates chemical
short range order\cite{PGehring2009}.
 
To understand the origin of the {\em shapes} of the diffuse scattering
we write the total intensity as a sum of Bragg and diffuse scattering,
and expand the latter in powers of ${\bf \xi}_\iti={\bf
r}_\iti-{\bf R}_{0 \iti}$, which being proportional to the local
polarization is a small quantity.  The Bragg term is $\sum_{i,j}{\it
e}^{{\bf iQ} \cdot ({\bf R}_{0i}-{\bf R}_{0j})}$, and the lowest order
diffuse scattering term is:
\begin{equation}
\label{eq:Idiff}
I_{diff}(\bf G,q) = <\sum_{\it i,j} |{\bf G+q}|^2 ({\bf \xi}_\iti
\cdot \hat\alpha)({\bf \xi}_\itj \cdot \hat\alpha)\cos({\bf Q} \cdot
({\bf R}_{0\iti}-{\bf R}_{0\itj}))>
\end{equation}
where {\bf q}={\bf Q-G} and {\bf G} is the Bragg spot. $\hat\alpha$ is
the unit vector along ${\bf G+q}$. The summation of pairs of atoms is
equivalently re-written as one over inter-atomic distances $\bf
{R}=\bf{R}_{0\iti}-\bf{R}_{0\itj}$, so that Eq.~\ref{eq:Idiff} becomes
the Fourier transform of projections of real space
displacement-displacement auto-correlation tensor ${\tensor C} ({\bf
R})$ along $\alpha$:
\begin{equation}
\label{eq:ftcorr}
I_{diff}(\bf G,q) = < \sum_{\bf R} (\alpha \cdot {\tensor C}
  ({\bf R}) \cdot \alpha) |{\bf Q}|^2 \cos({\bf Q} \cdot
{\bf R})>,
\end{equation}
%Then, we can write
%\begin{equation}
%\label{eq:forcorr}
%I_{diff}(\bf G,q) \approx < {\hat\alpha}_{\bf G} \cdot {\tilde {\tensor C} ({\bf q})} \cdot
%{\hat\alpha}_{\bf G} >,
%\end{equation}
where ${\tensor C} ({\bf R})$, given, in symmetric form is
\begin{equation}
\label{eq:realcorr}
C_{\alpha\beta} ({\bf R}) = 
\frac{1}{4} \sum_i (\xi_{i\alpha} \xi_{i\pm{\bf R}\beta} + \xi_{i\beta} \xi_{i\pm{\bf R}\alpha}),
\end{equation}
and the ``$\pm$" notation implies summation over both signs.

For small $|{\bf q}|$, we approximate ${\bf G+q}$ by ${\bf G}$ and
$\hat\alpha$ by the unit vector ${\hat\alpha}_{\bf G}$.  Fig 3. shows
the real space correlations in the relaxor phase (180K, 18GPa) with
projections along $\hat\alpha_{\bf G}$=[100] and [110] summed over the
'z' direction as well as their Fourier-transforms.  The diffuse
scattering butterfly and rod shapes come from $\alpha_{\bf G}$=[100]
and $\alpha_{\bf G}$=[110] projections, respectively, of the COR-COR
real-space correlations.  The CDR-CDR regions have strong correlation
along the direction perpendicular to the radial direction, leading to
the radial diffuse scattering in the Fourier space.  In 3D (not shown)
the [110] (as well as [111]) real space projections appear as
ellipsoids, while the [100] correlations appear as discs.
%Note
%that the distinctions between the COR and CDR 'Pb' atoms is not too
%stringent, especially at the COR-CDR boundaries.  This maybe the
%reason why we still see weak contributions to the butterfly and rod
%shapes from the CDR-CDR correlations. 

We conclude that PSN shows ferroelectric $\rightarrow$ relaxor
$\rightarrow$ paraelectric phase transition with increase in
temperature and/or pressure in our experiments.  The relaxor phase in
the experiments is characterized by butterfly and rod shaped diffuse
scattering.

Simulations further reveal that the anisotropic correlations from
COR-COR 'Pb' atoms which only have orientational degrees of freedom,
give rise to the
%BB change this:   "relaxor phase" to 
characteristic relaxor-phase diffuse scattering, rather than the
effect of strain or artificial atomic shifts or domain boundaries.
Polarization would be accompanied by local strain inhomogeneities,
that would cause additional contribution to the diffuse
scattering~\cite{Vakhrushev}, but are a {\em secondary} effect.  The
radial diffuse scattering is identified as coming from local
concentration fluctuations at the B-site, which reflect in the Pb
displacements from corresponding fluctuations in local electric-fields
on Pb. Coulomb energy is minimized when the local polarization aligns
parallel to the local field.  Our result suggests that by engineering
the shapes and relative positions of the COR regions, one could
possibly control the anisotropy in the COR-COR correlations, leading
to the design of new relaxor materials with superior electromechanical
properties.

An increase in pressure reduces the ferroelectric (free energy) well
depth and therefore the ferroelectric Pb-displacements in COR.  This
reduces PNR-PNR correlations (and not their sizes) and hence the
butterfly and rod shaped diffuse scattering.  In the ferroelectric
phase, the whole system, including the disordered matrix, has a
uniform polarization along $<$111$>$ directions.  This enhancement of
the local polarization reduces the effect of correlated
Pb-displacements in COR-COR pairs, greatly diminishing the
characteristic butterfly- and rod- shaped diffuse scattering features.
With increase in temperature, thermal fluctuations of the polarization
decrease the magnitude of the COR polarization, thereby decreasing the
COR-COR correlations.  Above $T_d$, only the CDR-CDR radial
contributions remain.

%Inhomogeneous strain may enhance the intensity of diffuse scattering, 
%but we have shown that they are not necessary to reproduce the experimentally 
%observed diffuse scattering shapes.  Rather, the strongest anisotropic 
%local polar-correlations derive from PNR-PNR correlations (spatially
%$\sim$ COR-COR correlations).  

\begin{acknowledgements}

The authors thank Stephen Gramsch and Eugene Venturini for their
useful discussions. This work was sponsored by the Office of Naval
Research under Grants No. N00014-07-1-0451 and N00014-02-1-0506; the
Carnegie/Department of Energy Alliance Center (CDAC,
DE-FC03-03NA00144). Use of the Advanced Photon Source was supported by
the U. S. Department of Energy under Contract No. DE-AC02-06CH11357.

\end{acknowledgements}

\bibliography{PSN9}

\begin{thebibliography}{23}
\bibitem[{Not()}]{Notes}
\bibinfo{note}{Simulations at each pressure-temperature were performed on a $40
  \times 40 \times 40$ supercell that contained 32000 atoms for about 300~ps
  with a time step of 0.6~fs using a Nose thermostat. The barostat only coupled
  to the homogeneous strain. Intensitites were computed after an equilibration
  time of 150~ps}.
\end{thebibliography}

\end{document}